\documentclass[pdflatex,sn-mathphys-num]{sn-jnl}

\usepackage{graphicx}%
\usepackage{multirow}%
\usepackage{amsmath,amssymb,amsfonts}%
\usepackage{amsthm}%
\usepackage{mathrsfs}%
\usepackage[title]{appendix}%
\usepackage{xcolor}%
\usepackage{textcomp}%
\usepackage{manyfoot}%
\usepackage{booktabs}%
\usepackage{algorithm}%
\usepackage{algorithmicx}%
\usepackage{algpseudocode}%
\usepackage{listings}%

\theoremstyle{thmstyleone}%
\theoremstyle{thmstyletwo}%

\theoremstyle{thmstylethree}%

\begin{document}

\title[Quantum Monte Carlo Simulations for predicting electron-positron pair production via the linear Breit-Wheeler process]{Quantum Monte Carlo Simulations for predicting electron-positron pair production via the linear Breit-Wheeler process}


\author*[1]{\fnm{Lucas Ivan} \sur{I\~nigo Gamiz}}\email{lucas.inigo.gamiz@tecnico.ulisboa.pt}

\author[1]{\fnm{Óscar} \sur{Amaro}}\email{oscar.amaro@tecnico.ulisboa.pt}

\author[2]{\fnm{Efstratios} \sur{Koukoutsis}}\email{stkoukoutsis@mail.ntua.gr}

\author*[1]{\fnm{Marija} \sur{Vranić}}\email{marija.vranic@tecnico.ulisboa.pt}

\affil*[1]{\orgdiv{GoLP/Instituto de Plasmas e Fusão Nuclear}, \orgname{Instituto Superior Tecnico, Universidade de Lisboa}, \orgaddress{\street{Av. Rovisco Pais 1}, \city{Lisboa}, \postcode{1049-001}, \country{Portugal}}}

\affil[2]{\orgdiv{School of Electrical and Computer Engineering}, \orgname{National Technical University of Athens}, \orgaddress{\street{Zagrophou}, \postcode{15780}, \country{Greece}}}


\abstract{Quantum computing (QC) has the potential to revolutionise the future of scientific simulations. To harness the capabilities that QC offers, we can integrate it into hybrid quantum-classical simulations, which can boost the capabilities of supercomputing by leveraging quantum modules that offer speedups over classical counterparts. One example is quantum Monte Carlo integration, which is theorised to achieve a quadratic speedup over classical Monte Carlo, making it suitable for high-energy physics, strong-field QED, and multiple scientific and industrial applications.  In this paper, we demonstrate that quantum Monte Carlo can be used to predict the number of pairs created when two photon beams collide head-on, a problem relevant to high-energy physics and intense laser-matter interactions. The results from the quantum simulations demonstrate high accuracy relative to theoretical predictions. The accuracy of the simulations is only constrained by the approximations required to embed polynomials and to initialise the quantum state. We also demonstrate that our algorithm can be used in current quantum hardware, providing up to 90$~\%$ accuracy relative to theoretical predictions. Furthermore, we propose pathways towards integrations with classical simulation codes.}


\keywords{Quantum Monte Carlo, Strong Field QED, Linear Breit-Wheeler, Quantum Simulations}



\maketitle

\section{Introduction}\label{sec1}

To overcome the size and cost limitations of conventional radio-frequency accelerator technology, plasma-based accelerators have emerged as an alternative for future high-energy colliders as they can sustain acceleration gradients exceeding $1~\textrm{GeV/cm}$. This emerging technology is not a replacement for conventional accelerators but a complementary tool, with access to Petawatt-class laser facilities providing a crucial testbed for non-linear strong-field QED (SFQED) regimes. Such effects include electron-positron pair production via nonlinear Breit-Wheeler  \cite{breit_collision_1934-1} and Bethe-Heitler \cite{motz_pair_1969} processes, as radiation emission via nonlinear Compton Scattering \cite{motz_pair_1969} and bremsstrahlung \cite{koch_bremsstrahlung_1959}. To model these effects, particle-in-cell (PIC) codes have been enhanced with QED modules to sample the event rates of quantum field theory \cite{fonseca_osiris_2002,fedeli_picsar-qed_2022}. For example, in electron-positron pair production, at each timestep, the QED module within the simulation loop conducts Monte Carlo (MC) calculations to replicate the probabilistic nature of QED events. If the sample size is large, these MC simulations can be computationally intensive even for the most powerful supercomputers in the world. By improving this module, we can enable larger or more accurate simulations and open the door to applying our solutions to any other problem that uses MC techniques.
\\
\\
Recent developments in quantum algorithms showcase the potential to surpass the computational capabilities of classical computers. Researchers have been developing algorithms demonstrating speedups over classical routines \cite{shor_polynomial-time_1997, brassard_quantum_2002}. In particular, Brassard \textit{et. al.}, demonstrated that, using quantum computers, one can perform quadratically fewer samples than a classical MC simulation to achieve the same accuracy \cite{brassard_quantum_2002}. This work was then adapted to use fewer gates and become more accessible to the Noisy Intermediate-Scale Quantum (NISQ) era \cite{nakaji_faster_2020,suzuki_amplitude_2020,grinko_iterative_2021}, with applications in pricing financial derivatives \cite{rebentrost_quantum_2018,woerner_quantum_2019,stamatopoulos_option_2020} and high-energy physics \cite{agliardi_quantum_2022}. Indeed, quantum information theory has been identified as a key area of high-energy physics, offering novel ways to probe fundamental physics \cite{afik_quantum_2025}.
\\
\\
In this paper, we present a NISQ hardware-conscious quantum Monte Carlo (QMC) integration that leverages iterative quantum amplitude estimation (IQAE) \cite{grinko_iterative_2021}. We develop the first readily applicable module that leverages state-of-the-art algorithms for SFQED and extreme plasmas, a field relatively unexposed to quantum computing. We choose the linear Breit-Wheeler electron-positron pair production process as a test base, as it is of great interest to the SFQED and high-energy physics (HEP) communities, and is proposed to be measured in numerous experiments. Additionally, the cross-section/probability can be approximated by a low-order polynomial, making it convenient for use in quantum algorithms. As MC integration is a necessary routine in a plethora of fields, the work performed here can be applied in fields that require stochastic sampling, such as Biology, Chemistry, Physics, Engineering, and Finance \cite{sakata_fully_2020,allison_recent_2016,nikjoo_s_uehara_w_e_wilson_m_hoshi_d_t_goodhead_track_1998, hartin_monte_2025,stevens_monte-carlo_2023,jackel_monte-carlo_2002}.

\subsection{\label{sec:level2}Strong Field QED}

Quantum electrodynamics (QED) is one of the most thoroughly tested fields of physics. QED effects are ubiquitous in space and in extreme astrophysical scenarios or objects such as black holes, neutron stars, and gamma-ray bursts \cite{piran_physics_2005,timokhin_time-dependent_2010,ruffini_electronpositron_2010}, rendering them crucial for understanding these astrophysical phenomena. It is worth noting that here we are not talking about the perturbative regime of QED. The strong-field quantum electrodynamics denotes a situation where the electromagnetic background is so strong that all possible contributions become relevant and the standard expansion used in QED calculations is not valid anymore. We do not take direct measurement in this regime, and the theory that exists so far is either too ideal because of the chosen geometry or incomplete in other ways.
\\
The next generation of Petawatt-class laser facilities is nearing its deployment \cite{papadopoulos_first_2019,bromage_technology_2019}. These laser facilities can deliver intensities up to 10 Petawatts, enabling studies of intense laser-matter interactions. These lasers interacting with plasmas can exceed the Schwinger limit, thus allowing strong-field QED effects, such as electron-positron pair production and vacuum birefringence, to occur. This places an urgency on studying SFQED to have accurate expectations for laboratory experiments. In this direction, Numerous experimental schemes have been proposed to leverage SFQED to study electron acceleration \cite{shaw_microcoulomb_2021,babjak_direct_2024,babjak_direct_2024-1}, positron production and acceleration \cite{maslarova_radiation-dominated_2023,martinez_creation_2023,martinez_direct_2025}, QED showers and cascades \cite{mercuri-baron_impact_2021,pouyez_multiplicity_2024, pouyez_properties_2025,grismayer_seeded_2017}. 

\subsection{\label{sec:level3}Quantum Computing}

In recent years, the interest in quantum information and computing research has steadily increased. Quantum computing companies are demonstrating significant improvements in hardware and software, as well as accessibility to their machines. The roadmaps project the deployment of fault-tolerant machines by the early to mid-2030s, aiming to demonstrate quantum advantage. Notably, financial institutions have reported the benefits, risks, and potential monetary gains of adopting quantum computing \cite{auer_quantum_2024}.
A rapidly developing area of application is quantum simulations, which harness quantum computers and algorithms to simulate physical systems or events, offering significant speedups (exponential, polynomial, and quadratic) over their classical counterparts \cite{harrow_quantum_2009,coppersmith_approximate_2002,brassard_quantum_2002,childs_toward_2018}. Additionally, quantum computers naturally enable the simulation of quantum systems. Classical computers usually need approximations and ``tricks" to account for the inherent quantum uncertainty. These manifest themselves in discrete classical models for stochasticity, which often require large particle statistics and formidable computing resources.
\\
A quantum circuit is an assembly of a discrete set of components which describe computational procedures \cite{nielsen_quantum_2010}. The basic unit of information of a quantum computer is the \textit{qubit}, which is the quantum analogue of a classical bit composed of a ground state ($|0\rangle$) and an excited state $(|1\rangle)$. Unlike the classical bit, the qubit allows for superposition of the ground and excited states:
\begin{equation}
    |\psi\rangle = \alpha|0\rangle + \beta|1\rangle
\end{equation}
where $|\psi\rangle$ is the quantum state, $\alpha$ and $\beta$ are the quantum amplitudes with $|\alpha|^2 + |\beta|^2 = 1$. A classical bit can be in only one state at a time. Superposition allows the qubit to hold more information than a classical bit. For example, to encode a state vector of 40 qubits, one requires $2^{40} \approx 1\mathrm{TB}$ of classical memory, entering the realm of high-performance computing.
\\
To perform operations in a quantum circuit, quantum gates, or operators are needed. There is a small number of universal gates: the \textit{Pauli matrices}, which can perform bit flips, phase flips and both simultaneously:
\begin{equation}
    X = \left[
    \begin{matrix}
        0 & 1
        \\
        1 & 0
    \end{matrix}
    \right];
    Y = \left[
    \begin{matrix}
        0 & -i
        \\
        i & 0
    \end{matrix}
    \right];
    Z = \left[
    \begin{matrix}
        1 & 0
        \\
        0 & -1
    \end{matrix}
    \right];
\end{equation}
By adding the $X$ and $Z$ matrices and dividing by $\sqrt{2}$, we get the Hadamard Gate, which turns a qubit into a superposition of both the $|0\rangle$ and $|1\rangle$ states. Furthermore, one can combine these gates, perform controlled operations on them, and construct a quantum algorithm. 
\\
At the time of writing, multiple quantum hardware technologies exist to perform digital quantum simulations. Examples include superconductors, trapped ions, neutral atoms, and photonic quantum computers. Although the current roadmaps suggest that by the mid-2030s, fault-tolerant machines will be deployed, we still need to consider NISQ devices, with simple, low-depth circuits that can perform a task.
\\
Quantum algorithms to perform quantum simulations of Strong Field QED and high-energy physics were recently proposed. Hidalgo and Draper have derived an SFQED Hamiltonian to simulate a real-time nonlinear Breit-Wheeler pair production and performed quantum simulations of a null double slit experiment \cite{hidalgo_quantum_2024}. Amaro \textit{et al.} have simulated the stochastic cooling of an electron beam under the influence of a strong magnetic field using Variational Quantum Imaginary Time Evolution \cite{amaro_variational_2025}.  Draper \textit{et al.} have performed digital quantum simulations of photon polarisation flips interacting with intense waves \cite{draper_hamiltonian_2025-1}. This paper introduces the first quantum algorithm for Monte Carlo in SFQED, leveraging iterative quantum amplitude estimation and state preparation techniques compatible with current NISQ hardware. We compare the performance of this routine with classical Monte Carlo routines and also show its performance on a trapped-ion quantum computer.

\bmhead{The algorithm}

The quantum Monte Carlo routine can be summarised as a four-step process: preparing an initial quantum state representing the photon beam with a varying energy profile; embedding a function which yields the probabilities for a selected energy range; performing the quantum equivalent of stochastic sampling; and post-processing the information from the quantum simulation. 

\begin{figure}[h]
    \centering
    \includegraphics[width=0.95\linewidth]{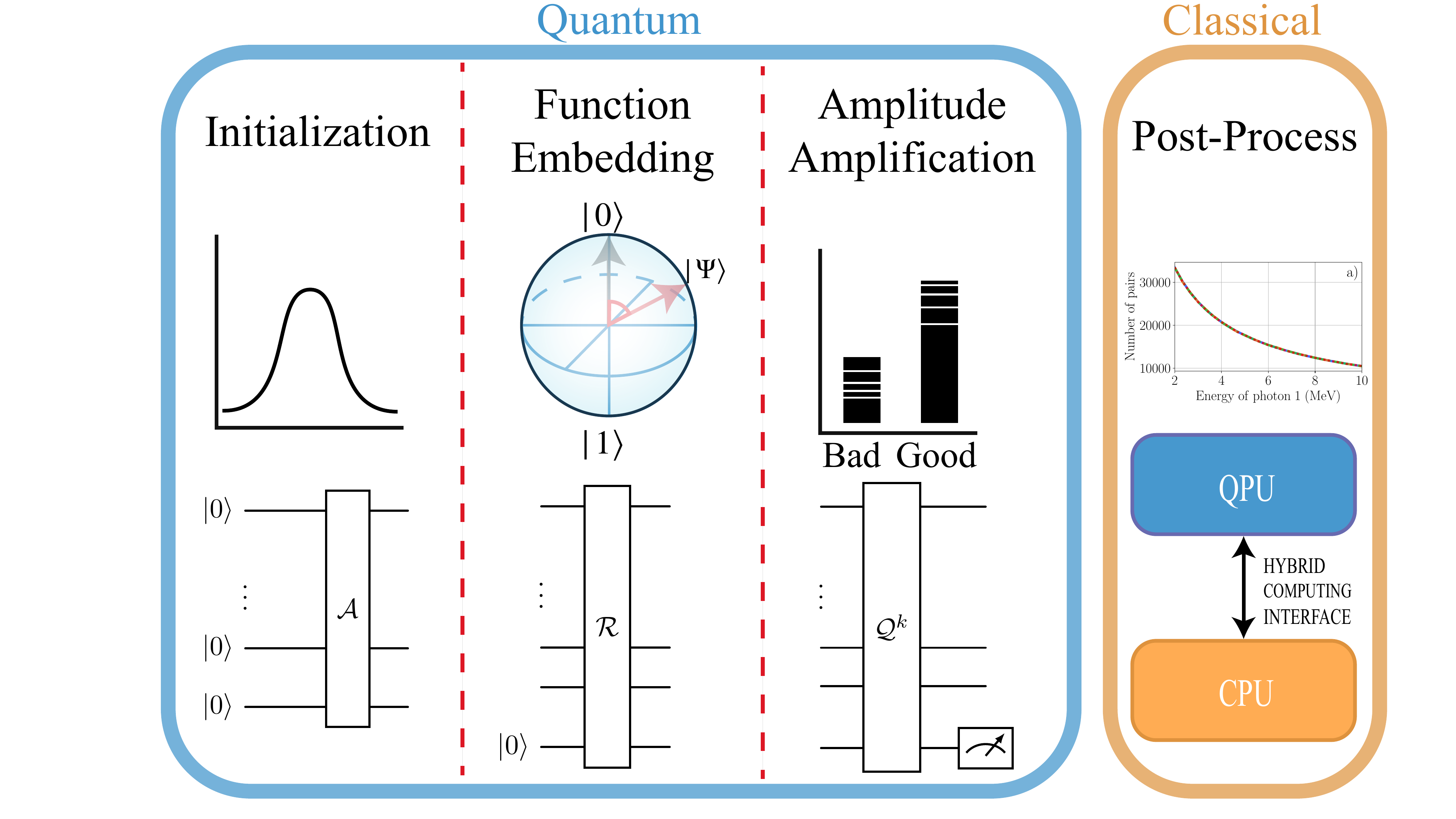}
    \caption{Cartoon showing the algorithm composed of its quantum and classical part. Within the quantum part, we have the initialisation of the probability distribution of the beam with a Gaussian energy distribution, the embedding of the probabilities to an ancillary function via controlled rotations; the amplitude amplificaction, of the ``good" state in the ancillary qubit. In the classical part, we have the post-processing part, where we the amplitudes of the probability in the ``good" state are read and interpreted to number of pairs produced via linear Breit-Wheeler.}
    \label{fig:algorithm pipeline}
\end{figure}

In figure ~\ref{fig:algorithm pipeline}, we show the four-step process for performing quantum Monte Carlo. In the initialisation box, the algorithm $\mathcal{A}$ initialises the quantum state, representing a probability distribution corresponding to one of the particle beams. Then, the probabilities can be embedded into an ancillary (or also called auxiliary) qubit in a series of controlled rotations via the algorithm $\mathcal{R}$ (please refer to the Supplementary Material for more details). Then, to perform the sampling, we perform iterative quantum amplitude amplification (IQAE) \cite{grinko_iterative_2021}, which amplifies the amplitude of a ``good" state in the ancillary qubit by performing algorithm $\mathcal{Q}, k$ times. The number of iterationsions is calculated within the iterative quantum amplitude estimation. Finally, from the measured results, we translate the expected amplitude from the simulation into the number of produced pairs. The steps for performing the simulation on a quantum computer are the same. For more detailed information on each step of the algorithm please check the Methods section and Supplementary Material.

\bmhead{Experimental Setup}
We test a head-on collision of $\gamma$ photons, with the geometry simplified so that all photons interact and have a nonzero probability of decaying into electron-positron pairs. In this case, we neglect further radiation emission and pair-producing mechanisms. One of the photon beams is monoenergetic, and the other has a Gaussian-like energy distribution. each photon beam has $10^{12}$ photons, and a photon density of $6\times10^{19}~\textrm{cm}^{-3}$.

\begin{figure}[h]
    \centering
    \includegraphics[width=\linewidth,keepaspectratio]{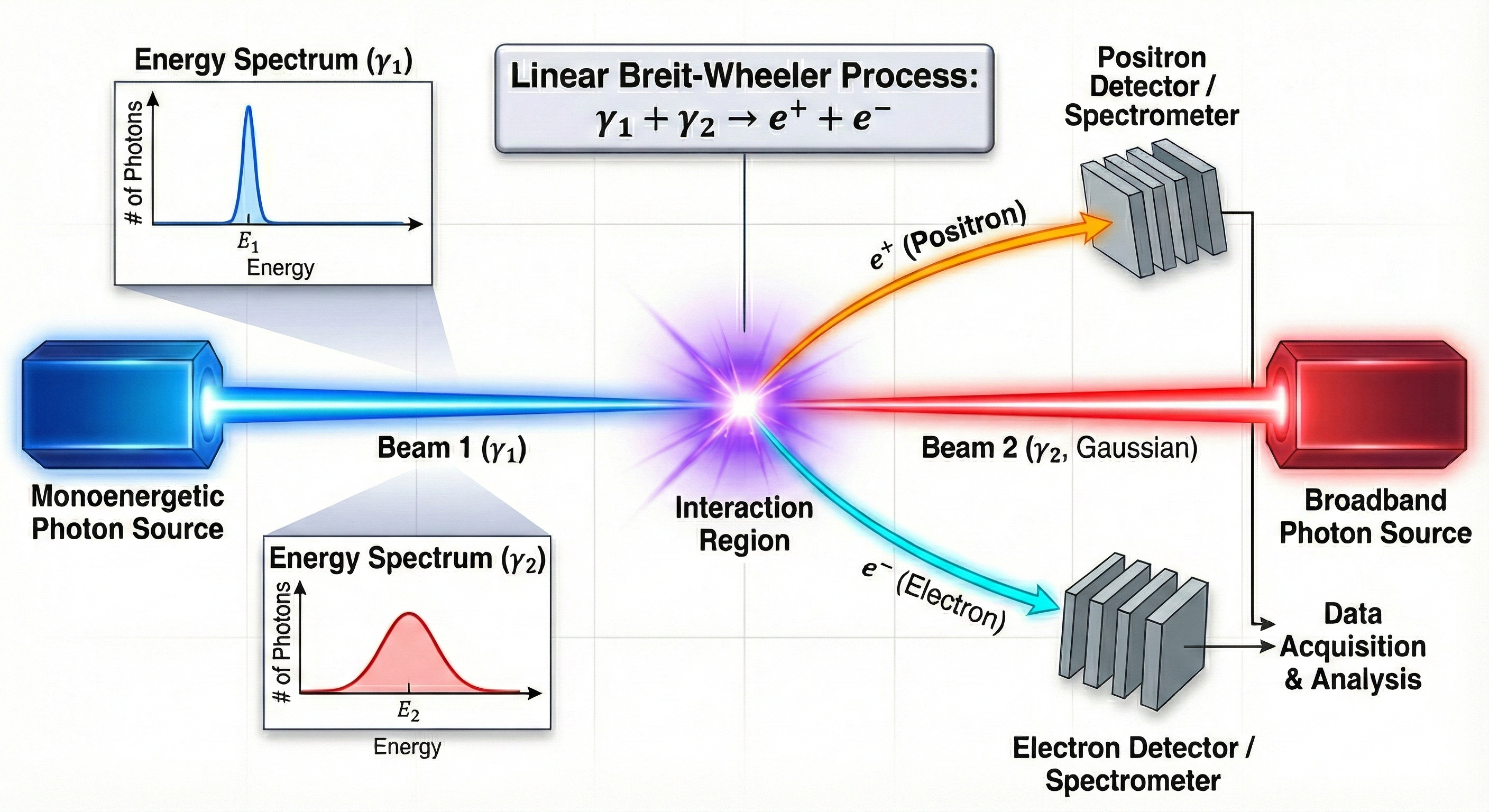}
    \caption{Cartoon showing an experimental setup of two beams colliding and creating electron-positron pairs via the linear Breit-Wheeler process.}
    \label{fig:experimental_setup}
\end{figure}

Figure~\ref{fig:experimental_setup} shows a representative schematic of our proposed setup to produce electron-positron pairs via the linear Breit-Wheeler process. Experimentalists have already used a similar schematic \cite{ribeyre_pair_2016}, but instead of having two head-on collisions, they have the collisions at an angle. Here, we simplify the geometry as much as possible to prepare the problem for use with current quantum algorithms and hardware.

\section{Results}\label{sec2}

In this section, we will simulate the number of pairs produced by two photon beams colliding head-on with each other. We begin by initialising the photon distribution, embedding the probabilities as ancillary-controlled rotations, performing Iterative Quantum Amplitude Estimation, and then doing post-processing to obtain the estimated number of pairs produced. We will investigate the robustness of the method and the accuracy relative to the theoretical expectation value, and compare the simulations by varying the monoenergetic beam energy, the Gaussian distribution width, the number of qubits, and the distribution skewness.

\bmhead{Comparison with ideal simulators}

\begin{figure}[h]
    \centering
    \includegraphics[width=\linewidth,keepaspectratio]{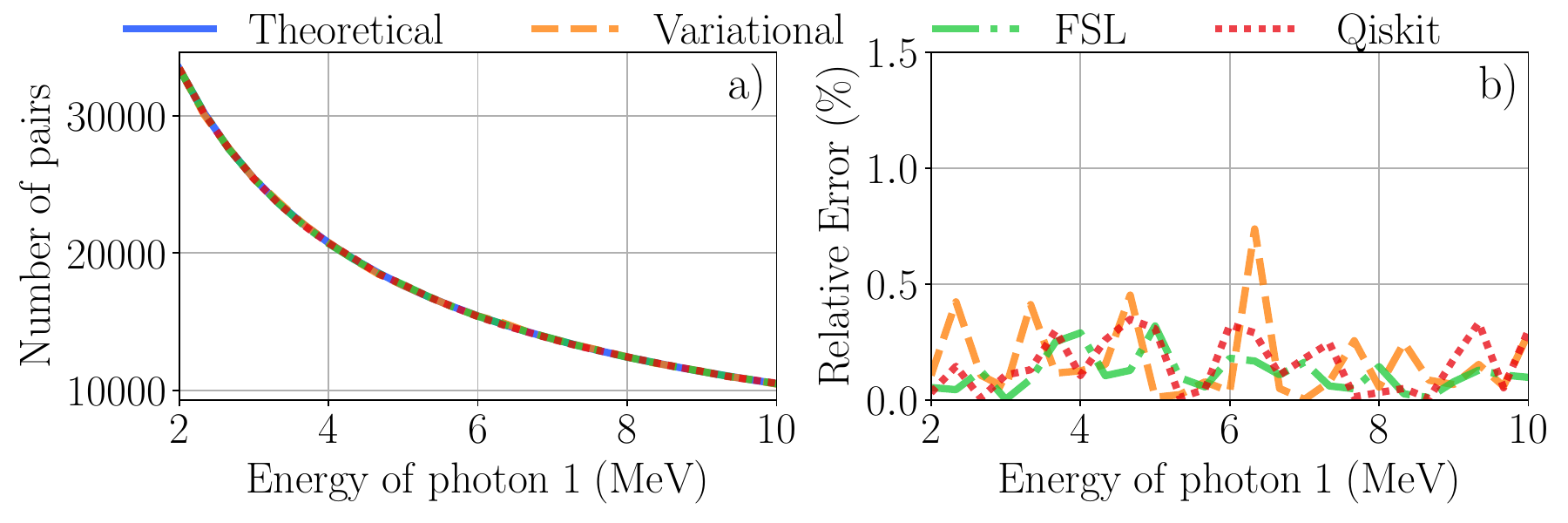}
    \caption{Panel a): Comparison between the theoretically predicted number of pairs (solid blue) and different state initialisation methods using a variational approach (spaced-dashed orange), Fourier Series Loader (FSL, dot-dashed green), and the initialisation method from Qiskit (dotted red). Panel b) comparison of the relative error with the different initialisation methods. The energy of the mono-energetic beam is from $2$ to $10~\textrm{MeV}$.}
    \label{fig:varying_energy_mono}
\end{figure}

Figure~\ref{fig:varying_energy_mono} shows the comparison between the number of pairs predicted theoretically and with the QMCI algorithm using different state preparation algorithms. The energy of the monoenergetic beam was varied whilst the parameters of the Gaussian beam were maintained the same, with its mean energy at $6~\textrm{MeV}$, and spread $\sigma = 1$.  One can observe from panel a) that the QMCI algorithm is indistinguishable from the theoretical results. As the beam energy increases, the probability of pair production decreases. Furthermore, we can observe that the QMCI algorithms, in general, are in excellent agreement with the theoretical predictions. In panel b), the relative errors for each method are shown. The relative error is consistently below 1.0~\%. The mean error for the Qiskit initialisation is 0.156~\%, for the FSL is 0.115~\%, and for the variational is 0.167~\%. For these simulations, we confirm a mean accuracy of 99.8~\% when compared to the analytical results. All state preparation methods achieve remarkable accuracy; however, only the variational and FSL methods can be used on quantum hardware, as the Qiskit native initialisation method uses operations that are not supported on quantum computers.

\begin{figure}[h]
    \centering
    \includegraphics[width=\linewidth,keepaspectratio]{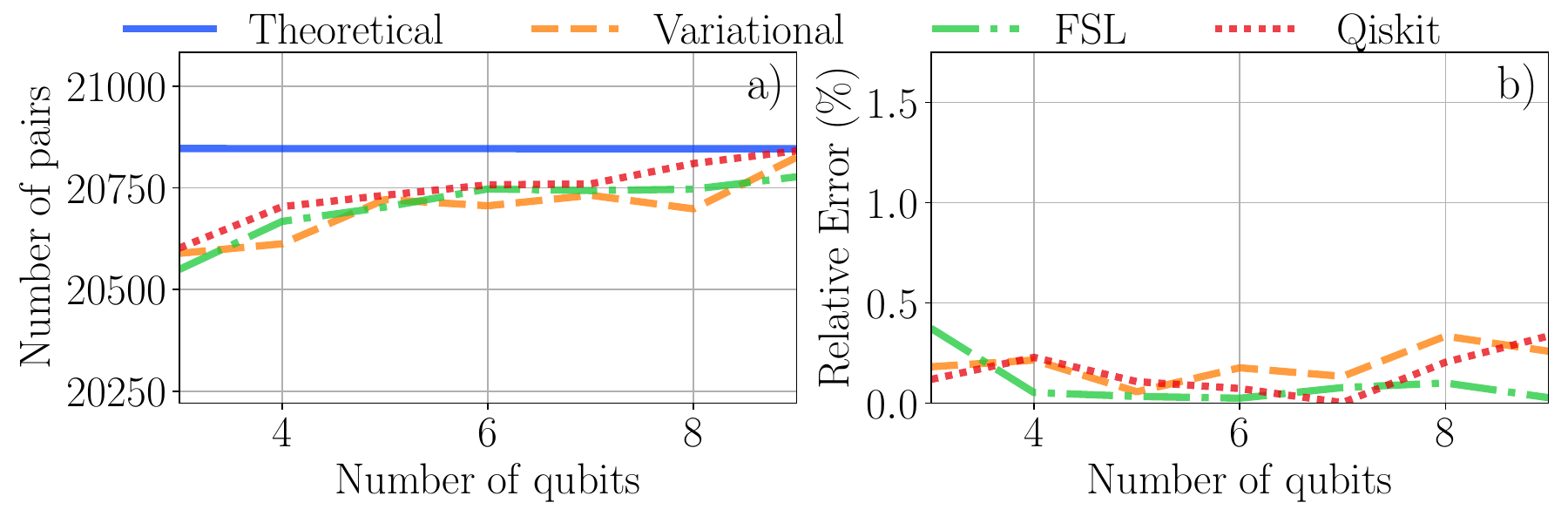}
    \caption{Panel a): Comparison between the theoretically predicted number of pairs (solid blue) and different state initialisation methods using a variational approach (spaced-dashed orange), Fourier Series Loader (FSL, dot-dashed green), and the initialise method from Qiskit (dotted red). Panel b) comparison of the relative error with the different initialisation methods. The number of qubits is varied.}
    \label{fig:varying_number_qubits}
\end{figure}

Figure~\ref{fig:varying_number_qubits} shows how the accuracy between each different state preparation is affected relative to the number of qubits. To precisely test accuracy, we also varied the discretisation of the theoretical results' initial distribution, yielding a non-constant value.  In this simulation, the energy of the monoenergetic beam is $4~\textrm{MeV}$, and the Gaussian beam has a mean energy of $6~\textrm{MeV}$ and a spread $\sigma = 1$. As the number of qubits increases, we get a better resolution, thus giving a more accurate result. With the FSL initialisation method, the error can be traced to the truncation order used. The number of qubits limits the truncation order. However, the variational approach differs. In this case, the higher the number of qubits, the more the ansatz tends to overfit, misrepresenting the initial Gaussian state. If we were to use more qubits, the initial state would be overfit.

\begin{figure}[h]
    \centering
    \includegraphics[width=\linewidth,keepaspectratio]{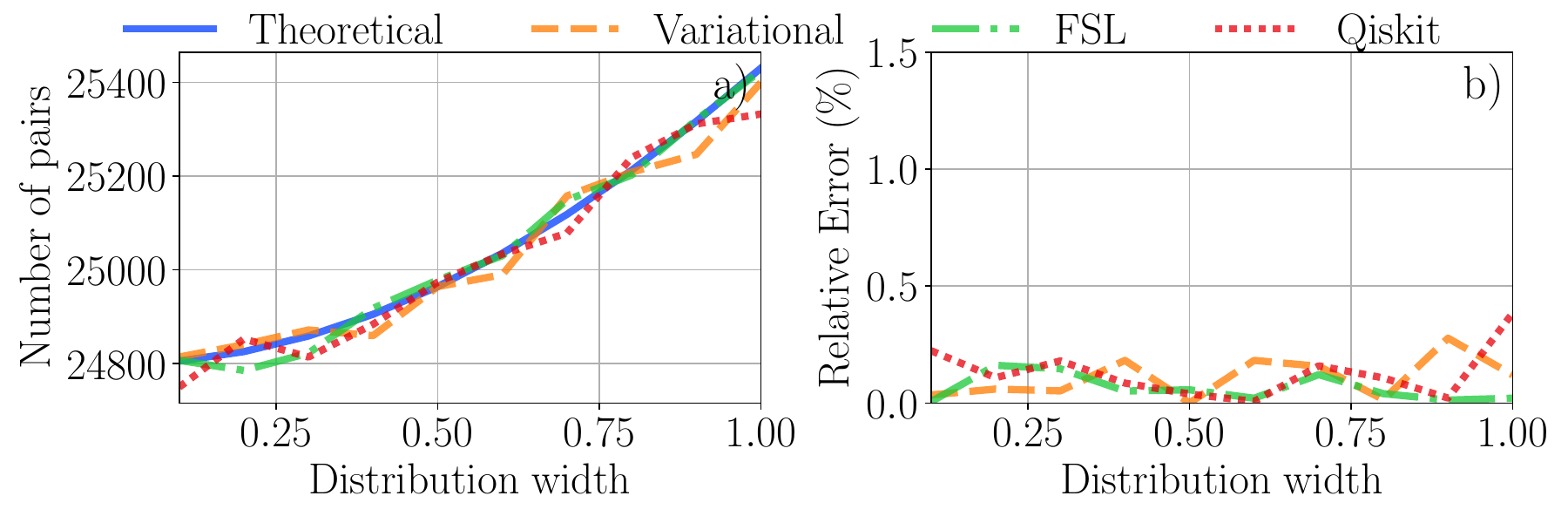}
    \caption{Panel a): Comparison between the theoretically predicted number of pairs (solid blue) and different state initialisation methods using a variational approach (spaced-dashed orange), Fourier Series Loader (FSL, dot-dashed green), and the initialise method from Qiskit (dotted red). Panel b) comparison of the relative error with the different initialisation methods. The spread of the distribution is varied.}
    \label{fig:varying_sigma}
\end{figure}

Figure~\ref{fig:varying_sigma} shows how the spread of the distribution affects the predicted number of pairs using QMCI. The simulations consistently show over $99\%$ accuracy compared to the theoretical predictions across all initialisation schemes. All the schemes have a similar error rate; a key component would be a discretisation which is fine enough to resolve different spreads of the distribution. Using fewer qubits would yield a higher discrepancy.

\begin{figure}[h]
    \centering
    \includegraphics[width=\linewidth,keepaspectratio]{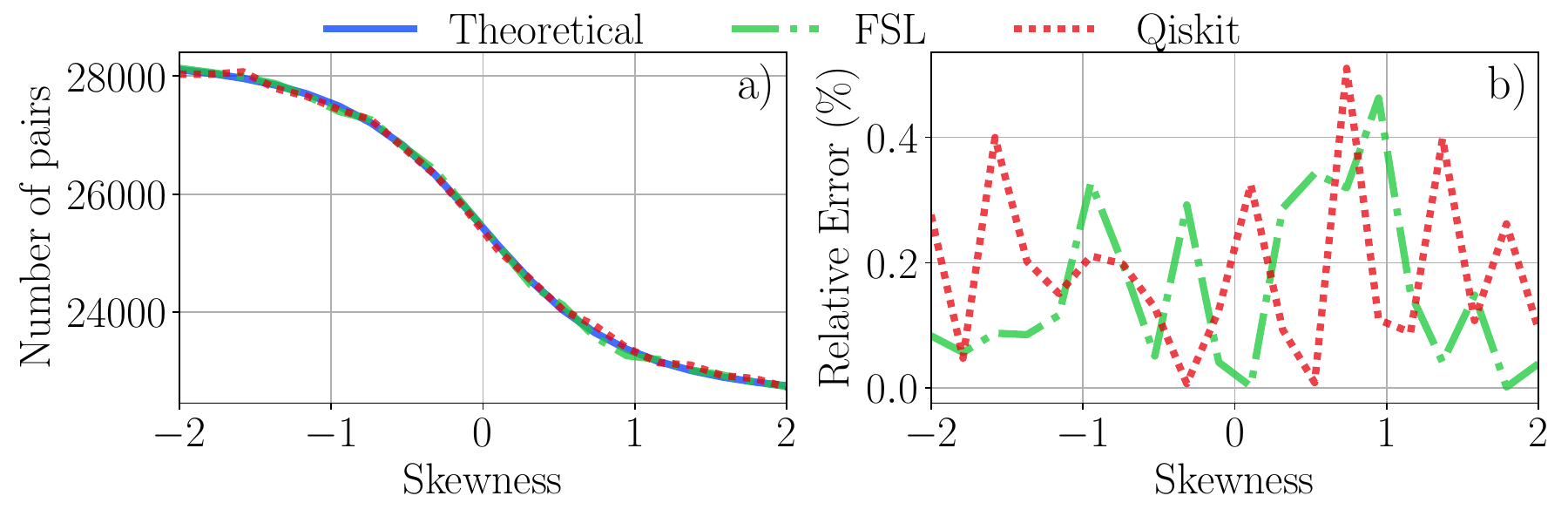}
    \caption{Panel a): Comparison between the theoretically predicted number of pairs (solid blue) and different state initialisation methods using the Fourier Series Loader (FSL, dot-dashed green), and the initialise method from Qiskit (dotted red). Panel b) comparison of the relative error with the different initialisation methods. The skewness is changed to account for different distributions.}
    \label{fig:skewed}
\end{figure}

Figure~\ref{fig:skewed} shows the QMCI algorithm performance with a skewed Gaussian distribution. In this case, the energy of photons in the monoenergetic beam is $4~\textrm{MeV}$,  the spread $\sigma = 1$, and the skewness is varied from positive to negative skewness $[-2,2]$. The distribution has a mean energy of $6~\textrm{MeV}$, and we use six qubits for this simulation. It is noticeable that the FSL performs worse on negatively skewed distributions than on positively skewed distributions. However, the accuracy remains consistently above 99~\%. The variational approach we used does not allow for skewed distributions. See the Supplementary Material for more details.

\subsection{Comparison Classical vs Quantum Monte Carlo}

We also performed classical MC simulations to compare with our previously discussed results. Similarly, we have our Breit-Wheeler pair production probability $p(X) \in [0,1]$ for a single interaction. Our value of interest is the expectation value of the probability of pair production:
\begin{equation}
    \mu = \mathbb{E}_{X}[p(X)],
\end{equation}
where the expected number of pairs for the number of interacting particles is $N_\textrm{pairs} = N_\textrm{int}\mu$. 

We can compare our results using the same oracle queries in the QMC and samples used in the MC simulations. The number of oracle calls is defined as the number of times the Grover operator is applied $(k)$ times the number of shots per simulation run. Within IQAE, The \verb|FindNextK| algorithm (outlined in \cite{grinko_iterative_2021}), the powers $k$ go as $k_j = 4k_i + 2$. Consequently, during iterative amplitude estimation, this $k$ is computed and then applied to the Grover operators. We compared the average error within each simulation to that of classical MC with the same number of queries per simulation.

\begin{table}[h]
    \centering
    \begin{tabular}{|c|c|}
        \hline
         Simulation Type & Mean \% Error vs Analytic \\
         \hline
         Classical & $0.191$\\
         \hline
         Qiskit  & $0.156$ \\
         \hline
         Variational  & $0.167$\\
        \hline
         FSL & $0.115$
         \\
         \hline
    \end{tabular}
    \caption{Comparison between classical and Quantum Monte Carlo for predicting the amount of electron-positron pairs produced via linear Breit-Wheeler where each simulation had the same number of queries/samples of 6144.}
    \label{tab:comparison_cmc_qmc}
\end{table}

As shown in Table~\ref{tab:comparison_cmc_qmc}, the QMC routine consistently indicates an improvement over the classical MC method in terms of accuracy. We also tested the accuracy between classical and quantum MC for varying values of $\epsilon$, which yielded results comparable to those shown here and are omitted for simplicity. 

\bmhead{Comparison with trapped-Ion QPUs}

We have adapted the results from our simulations to apply to a quantum computer. For this simulation, we compare the idealised simulation available from the cloud service provider with simulations performed on IonQ's Forte Enterprise quantum computer. We vary the energy of the monoenergetic beam whilst keeping the Gaussian beam energy distribution unchanged. We decreased the number of qubits in the working register to minimise cost. 

\begin{figure}[h]
    \centering
    \includegraphics[width=\linewidth,keepaspectratio]{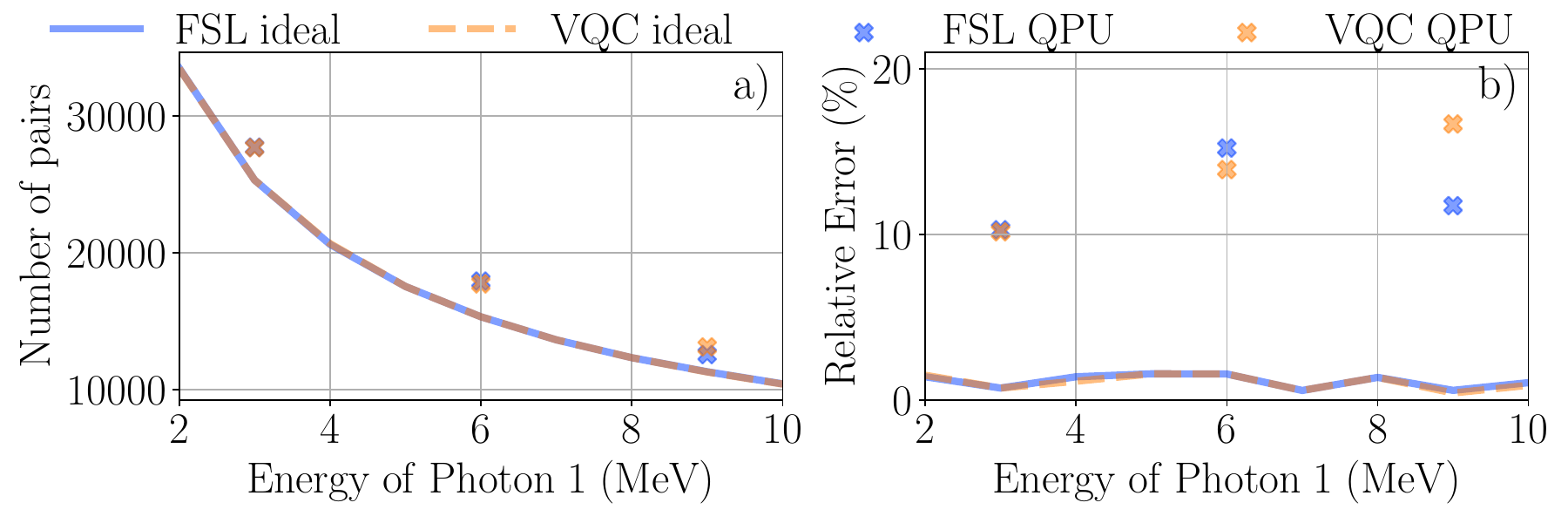}
    \caption{Panel a): Comparison between the different state initialisation methods using a variational approach (dashed orange), Fourier Series Loader (FSL, solid blue), and the simulations from the quantum computer with the respective initialisation methods, blue cross for FLS and orange cross for VQC. Panel b) comparison of the relative error with the different initialisation methods and the QPU simulations The energy of the mono-energetic beam is from $2$ to $10~\textrm{MeV}$}
    \label{fig:varying_energy_mono_QPU}
\end{figure}

In Figure \ref{fig:varying_energy_mono_QPU}, we can compare the performance of the simulations with the results from the QPUs. We observe that the algorithm performs well even with fewer qubits, and the results from the quantum computer can be over $90~\%$ accurate without error correction. The trends of the simulations from the quantum computers follow a similar behaviour to the varying energy of the monoenergetic beam seen in the ideal simulations, showing that the algorithm is effectively amplifying the ``good" state in the ancillary qubit. By using an error-correcting algorithm, using more qubits, and increasing the number of shots, we can potentially improve accuracy.

\section{Conclusion}\label{sec13}

We have constructed an algorithm to perform quantum Monte Carlo simulations to predict the electron-positron pairs produced via the linear Breit-Wheeler channel, readily applicable to NISQ-era devices. Here, we present the entire algorithm, which is built upon other algorithms to initialise a distribution function, apply a function and rotations to an ancillary qubit, perform amplitude estimation, and proceed with post-processing to accurately obtain the expected number of pairs produced by two colliding high-energy photons. We assess the robustness of the algorithm by comparing the theoretically predicted number of pairs with quantum simulations, varying the energy of a monoenergetic photon beam while keeping the beam with a Gaussian distribution fixed. We have also varied the width of the Gaussian distribution, the number of qubits, and the skewness of the Gaussian distribution. The idealised results demonstrate remarkable accuracy, exceeding that of classical Monte Carlo. We have also performed simulations on IonQ's Forte-Enterprise QPU, demonstrating the feasibility of our algorithms on current NISQ hardware and achieving a mean accuracy of $87~\%$ relative to theoretical predictions. 

This study arrives as quantum computers are becoming increasingly accessible and multiple national and international initiatives are advancing towards hybrid classical-quantum codes. The results shown demonstrate that a speedup is indeed achieved, limited by approximations to the initialisation function and the embedded polynomial. Furthermore, it suggests that these studies can be conducted on current NISQ devices, albeit with limited qubit counts, thereby reducing the accuracy of the routine. 

Although the work is restricted to a univariate problem, we have identified a possible avenue for introducing multivariable problems, such as varying energy profiles between both particle species, spin, polarisation, etc. Thus, we can extend possible scenarios in which quantum offers an advantage over classical MC. Consequently, our next steps would be to expand the algorithm to multiple variables. Our subsequent studies will focus on adapting QMC to SFQED modules within classical, enabling a meaningful integration between HPC and QPUs, and preparing for future collider experiments.

\section{Methods}\label{sec11}

\bmhead{Quantum Simulation using Quantum Emulators}

We approximate the Breit-Wheeler probability/cross-section ($\sigma_{\gamma \gamma}(s)$) over the relevant energy interval using a second-degree polynomial $\tilde{p}(x) = a_0 + a_1 x + a_2 x^2$. To ensure that $\tilde{p}(x) \in [0,1]$, we rescale and shift the coefficients using equation \textbf{See supplementary material} \ref{eqn:scale}. The polynomial is then embedded via ancilla rotation $R_y(2\arcsin(\sqrt{\tilde{p}(x)}))$ conditioned on the work register. We implement this using Qiskit's \verb|PolynomialPauliRotations|, which decomposes the polynomial into a sequence of multi-controlled rotations. The mean error between the original polynomial and this technique is 0.69\%. The classical Monte Carlo simulations are set to have the same sample size as the quantum queries ($\text{number of shots} \times \text{Grover operators}$), and are calculated within the IQAE loop. The classical Monte Carlo method is performed, for simplicity, by adding an arbitrary number of samples drawn from the Gaussian distribution. They were performed using \verb|random| module of the \verb|NumPy| Python framework. 
Second results: the photons of the monoenergetic beam have an energy of $4~\textrm{MeV}$, and the Gaussian beam energy spread is varied from $0.1$ to $1.0$ with its mean at $6~\textrm{MeV}$.

\bmhead{Quantum Simulation using Quantum Computers}

We used IonQ's Forte-enterprise quantum computer available at IonQ's Quantum Cloud. The quantum processor has 36 physical Qubits with an all-to-all connectivity topology, a one-qubit median gate error of $0.02\%$, and a median two-qubit error of $0.450\%$. The quantum circuit was prepared using Qiskit and the quantum Monte Carlo integration module we have developed. The quantum circuit is transpiled specifically for the quantum computer's backend. The quantum circuit has three qubits in its work register, which is initialised using a parametric circuit and a Fourier series, as described in the Supplementary material. The embedding of the polynomial is done in the same way as described before, and in the supplementary material section. The results are downloaded in a JSON format and then post-processed locally. Six simulations were performed with 6144 shots. The simulations consisted in varying the energy as 3, 6 and 9 MeV of the monoenergetic beam, whilst keeping the other beam with the same energy distribution.

\backmatter

\bmhead{Supplementary information}

\bmhead{Linear Breit-Wheeler Pair Production}

When two energetic photons interact, they can decay into an electron-positron pair \cite{ritus_quantum_1985}. The linear Breit-Wheeler (LBW) process is a first-order perturbative QED process which describes the electron-positron pair decay from two interacting photons as $\gamma' + \gamma \rightarrow e^+ + e^- $. This is then followed by multiphoton processes \cite{ribeyre_pair_2016}. Breit-Wheeler pairs are difficult to detect in a laboratory setting, as other pair-production mechanisms can occur, such as the Bethe-Heitler \cite{motz_pair_1969} and the Trident pair-production mechanisms \cite{vodopiyanov_effect_2015}, which can overshadow the Breit-Wheeler process.

This process was observed experimentally. In the SLAC experiment, an intense laser colliding with a 46.6~$\textrm{GeV}$ electron beam, produces gamma photons which then interact with the laser field, creating electron-positron pairs \cite{burke_positron_1997}, obtaining the first experimental evidence of antimatter generated using lasers (albeit at a very low yield, approximately $100$s of positrons over the entire campaign). 
Assuming that electrons and positrons are produced at rest in the centre-of-mass frame of reference, we can write the threshold condition
\begin{equation}
    E_{\gamma_1}E_{\gamma_2} = 2m^2_ec^4 / (1 - \cos(\phi))
    \label{eqn:threshold}
\end{equation}
Where $E_{\gamma1}, E_{\gamma2}$ correspond to the photon energies, $m_e$ is the electron mass, $c$ is the speed of light, and $\phi$ is the collision angle between the photons. For an ideal collision geometry, a head-on collision at $\phi = \pi$ should be used to maximise the pair-production probability. 

To calculate the amount of electron-positron pairs produced, we can use the following cross-section \cite{ribeyre_pair_2016}:
\begin{equation}
    \sigma_{\gamma\gamma}(s) = \frac{\pi}{2}r_e^2 (1 - \beta^2) \left[ - 2 \beta \left( 2 - \beta^2\right)  + \left(3 - \beta^4\right)\ln \frac{1 + \beta}{1 - \beta}\right]
    \label{eq:cross_section}
\end{equation}
Where $s = E_{\gamma_1}E_{\gamma2} (1 - \cos(\phi)) / (2 m_e^2 c^4)$, $\beta = \sqrt{1 - 1 / s}$ and $r_e$ is the electron radius. The threshold energy for pair production for LBW should be above $0.5~\mathrm{MeV}$. The probability of pair production can be calculated as:
\begin{equation}
    p(E_{\gamma_1},E_{\gamma_2}) = 1 - e^{\sigma_{\gamma_1\gamma_2} N_{\gamma_2}/ A }
\end{equation}
Where $N_{\gamma_2}$ is the number of photons of beam 2 and $A$ is the interaction area.

\begin{figure}
    \centering
    \includegraphics[width=0.9\linewidth,keepaspectratio]{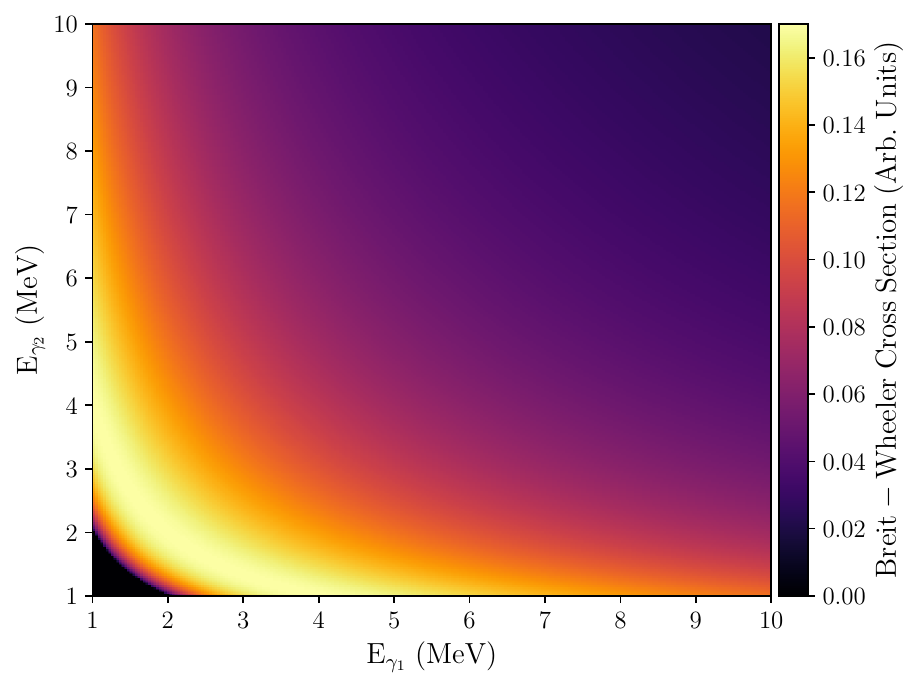}
    \caption{A colour map showing the cross section of the linear Breit-Wheeler pair-production process as a function of the centre-of-mass energy $s$ and a head-on collision. The colour map is the normalized Breit-Wheeler cross-section $\sigma_{\gamma \gamma}$}
    \label{fig:BW_cs}
\end{figure}

Figure \ref{fig:BW_cs} illustrates the dependence of the LBW cross-section on the energies of the two photon beams colliding head-on. The figure demonstrates that at high $s$, the cross-section decreases, thereby hindering the probability of producing electron-positron pairs. Consequently, one has to balance the energies between both to create the largest number of pairs.

\bmhead{Quantum Amplitude Estimation}

In the 2000s, Brassard \textit{et al.} \cite{brassard_quantum_2002} introduced an algorithm which expands beyond Grover's algorithm and allows for the amplification and estimation of a target state. They state that, by having an un-initialised $N-$qubit quantum circuit $|0^{\otimes N}\rangle$, we can initialize this with an algorithm 
$\mathcal{A}$ as 
\begin{equation}
\mathcal{A}|0^{\otimes N}\rangle \rightarrow \sum_{i=0}^{2^N-1}\sqrt{a_i}|i\rangle\label{eqn:algo_1}
\end{equation}
where $a_i$ is the amplitude corresponding to basis state $i$. We can also choose a basis in which our state looks like
\begin{equation}
\sum_{i=0}^{2^N-1}\sqrt{a_i}|i\rangle = \sqrt{1-p}|\Psi_0\rangle + \sqrt{p}|\Psi_1\rangle,
\end{equation}
Where $|\Psi_0\rangle$ and $|\Psi_1\rangle$ correspond to a new basis representation where the former is a \textit{bad} state and the latter is a \textit{good} state, and what we want is to amplify the amplitude of the good state and then measure it. This can be done using the Grover operators:
\begin{equation}
    \mathcal{Q} = \mathcal{A}\mathcal{G}_0\mathcal{A}^\dagger\mathcal{G}_1,
    \label{eqn:grover_ops}
\end{equation}
with:
\begin{align}
    \mathcal{G}_0 &= 2|\Psi_0\rangle\langle\Psi_0| - \mathrm{I},
    \\
    \mathcal{G}_1 &= \mathrm{I} - 2|\Psi_1\rangle\langle\Psi_1|,
\end{align}
Where $\mathrm{I}$ is the identity matrix, $\mathcal{G}_0$ is a rotation on the bad state, and $\mathcal{G}_1$ is a rotation on the good state. The operator $\mathcal{Q}$ is often referred to as a \textit{Grover iteration}. 
The measured integer is mapped to an angle $\theta_a = y \pi / M$, where $y = \lbrace 0,\dots,M-1 \rbrace$, $M=2^m$ where $M$ is the number of samples and $m$ is number of ancillary (helper) qubits, and the estimate is defined as $\tilde{a} = \sin^2(\tilde{\theta}_a)$. Consequently, the estimate of $a$ is then \cite{grinko_iterative_2021}
\begin{equation}
    |a - \tilde{a}| 
    \leq \frac{2\pi \sqrt{a (1 - a)}}{M} + \frac{\pi^2}{M^2}.
\end{equation}
QAE has been used to price financial instruments, such as derivatives, to conduct credit risk analysis, to optimise portfolios \cite{rebentrost_quantum_2018,woerner_quantum_2019,stamatopoulos_option_2020}, and to predict cross-sections in high-energy physics \cite{agliardi_quantum_2022}. There are a few more versions of quantum amplitude estimation that allow for faster, more accurate routine \cite{grinko_iterative_2021,lejarza_quantum_2023}. 

\bmhead{Iterative Quantum Amplitude Estimation}

The canonical QAE, although effective, can be computationally expensive, and depending on the problem, it may be inaccessible for current NISQ devices. Besides, it relies on quantum phase estimation, where it uses $m$ ancilla qubits to represent the final result and applies geometrically increasing powers of $\mathcal{Q}$ controlled by ancillas. Consequently, the reliance of the canonical QAE on quantum phase estimation is not ideal for current NISQ devices.
Recently, a faster and more accurate way of performing amplitude estimation was proposed by Grinko \textit{et al} \cite{grinko_iterative_2021}, based on the fact that 
\begin{equation}
    \mathcal{Q}^k \mathcal{A}|0^{\otimes n}\rangle|0\rangle = \cos((2k + 1) \theta_a) |\Psi_0\rangle|0\rangle
    + \sin((2k+1)\theta_a)|\Psi_1\rangle|1\rangle,
\end{equation}
and the probability of measuring $|1\rangle$ in the ancillary register is 
\begin{equation}
    \mathbb{P}[|1\rangle] = \sin^2\left((2k + 1) \theta_a\right),
\end{equation}
Which means that we measure the last qubit in $\mathcal{Q}^k \mathcal{A} |0^{\otimes N}\rangle|0\rangle$ for different powers of $k$. Consequently, we need to redefine the angle $\theta_a$ as $a = \sin^2(\theta_a)$. Where $a$ is then obtained as an input from a confidence interval $1 - a$, in the IQAE algorithm, the maximum number of applications of the Grover operator can be calculated as the number of times the Grover operator is applied, serving as a loose upper bound. It uses the Chernoff-Hoeffding bound to estimate sufficiently narrow confidence intervals from our initial conditions. The maximum number of shots, based on the confidence interval and accuracy $\epsilon$, is calculated as:
\begin{equation}
    N_\mathrm{max}(\epsilon, a) = \frac{32}{(1 - 2\sin(\pi / 14))^2}\log \left( \frac{2}{a} \log_2\left(\frac{\pi}{4\epsilon} \right)\right).
\end{equation}
However, as the algorithm iteratively checks whether we have satisfied our predetermined accuracy, we may not need that many shots.

\bmhead{Initialisation by variational methods}

Variational Algorithms have been proven helpful in the NISQ era due to their minimal circuit requirements and ease of use on current machines. They have proven beneficial in many fields of science, ranging from Quantum Chemistry \cite{peruzzo_variational_2014} and Finance \cite{wang_variational_2025,rebentrost_quantum_2024} to more recent applications in SFQED \cite{amaro_variational_2025}. Here, we adopt the ansatz proposed by Amaro et al., which is an ansatz composed of $\text{CNOT}-R_y$ in a ring structure. It has CNOTs to enforce symmetry, further reducing the computational requirements while retaining expressibility.

\begin{figure}[h!]
    \centering
    \includegraphics[width=\linewidth,keepaspectratio]{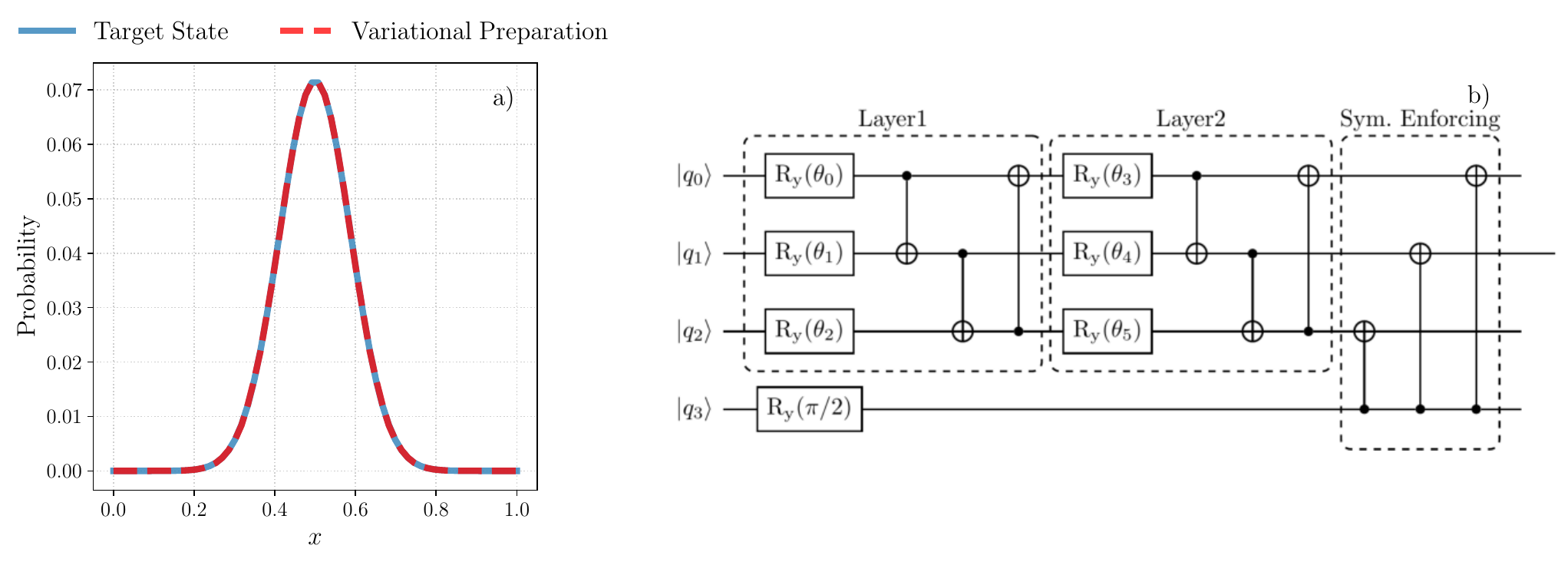}
    \caption{a) Comparison between the Gaussian target state and the results using a variational algorithm. b) A schematic of the quantum circuit used to produce the results in panel a).}
    \label{fig:variational_circuit}
\end{figure}
    
In Figure~\ref{fig:variational_circuit} panel a), we observe that, for a circuit with five qubits, a Gaussian is well represented. The cost to initialise this Gaussian is $O(nL) R_z$ gates, where $n$ is the number of qubits, $L$ is the number of layers, and $O(nL)$ CNOT gates. Although this is a cheap and simple method for embedding Gaussian distributions, it cannot be applied immediately to skewed distributions; it can only be applied to those that exhibit symmetry. Furthermore, the ansatz struggles when the working register exceeds six qubits.

\bmhead{Initialisation by Fourier series loading}

Although other options could potentially be less computationally expensive \cite{zylberman_efficient_2024,feniou_sparse_2024}, we opted for an ancilla-less approach, where it initially initialises a circuit that uses a truncated Fourier series \cite{moosa_linear-depth_2024}. The Fourier coefficients are calculated classically and then truncated to order $m$. Furthermore, there are two potential approaches to initialise a state with the Fourier Series Loading (FSL): using a set of cascading controlled rotations and using the Schmidt decomposition to obtain a unitary gate $U_c$. Consequently, the truncated Fourier is embedded into a register of size $m + 1$. Then an inverse Fourier transform is applied to the entire circuit, efficiently initialising an arbitrary state to the quantum circuit \cite{moosa_linear-depth_2024}. 
\\
\begin{figure}[h]
    \centering
    \includegraphics[width=0.9\linewidth,keepaspectratio]{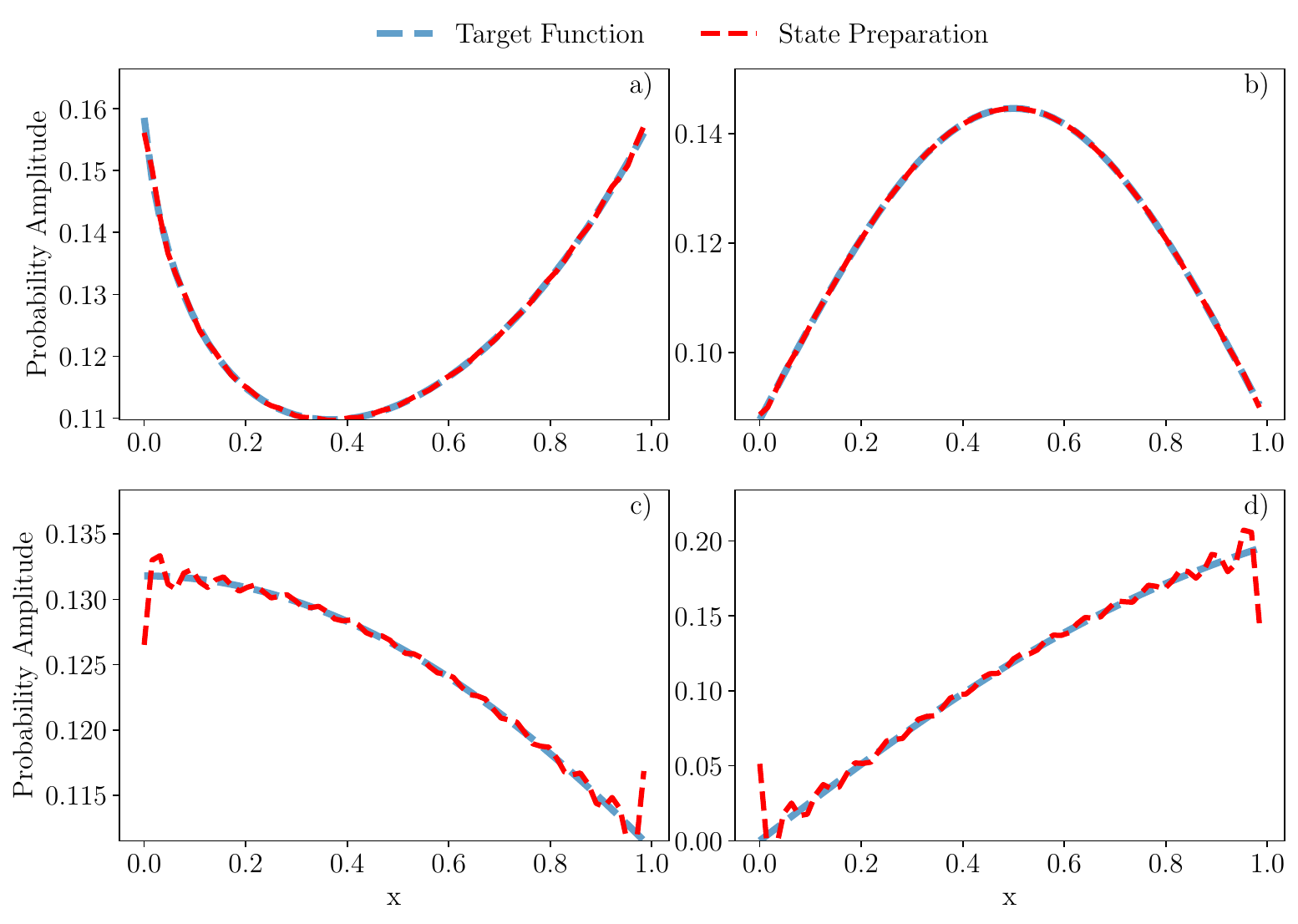}
    \caption{State initialisation using of $f(x) = \lbrace x^x, \frac{1}{\sqrt{4 \pi \sigma}}e^{\frac{- x^2}{2\sigma^2}},\text{sinc}(x),\tanh(x) \rbrace$ using the Fourier Series Loader method in a), b), c), and d) respectively. We use six qubits, with a truncation number $m=4$ to fit each function.}
    \label{fig:state_prep}
\end{figure}
\\
From the figure \ref{fig:state_prep}, we can see that the FSL effectively loads an arbitrary function to the state, which we can then use to perform amplitude estimation. The cost of using this method is $DN +2^{D(m+1) +1} - 1$ single gate operations, and $DN(N+1)/2 + 2^{D(m+1)+1} - 3D(m-1)-1$ two-qubit gate operations where $D$ is the dimension, and $m$ is a free parameter related to the truncation order.

\bmhead{Quantum Monte Carlo routine}

Having illustrated the core components of the quantum amplitude estimation, we can now unveil the workflow for the MC integration. In this section, we will expand on the introduction of QAE to include algorithms for preparing a quantum state, embedding a polynomial function, and performing amplitude amplification and estimation. 
The experimental setup consists of two counter-propagating particle beams. One of them is monoenergetic, while the other has an initial energy distribution. The two particle beams will collide head-on, maximising the interaction time and thus the pair production probability. 

We adopt the approach presented by Rebentrost~\textit{et al} \cite{rebentrost_quantum_2018}, and Woerner and Egger \cite{woerner_quantum_2019}, but instead of pricing derivatives, we adapt it to calculate the expected number of pairs produced. We begin by initialising a quantum state $|0^{\otimes N}\rangle$ to have an initial probability distribution corresponding to the particle energy. The initialisation looks like in equation~\ref{eqn:algo_1}:
\begin{equation}
    \mathcal{A}|0^{\otimes N}\rangle = \sum_{x = 0}^{2^N-1}\sqrt{p_x}|x\rangle,
\end{equation}
$N$ is the number of qubits, $p_i$ is the probability, and $x_i$ is the discretised energy grid. Later in the manuscript, we present two algorithms for initialisation.

For functions bounded between $[0,1]$, we can map the function into an ancillary qubit by performing rotations as:
\begin{equation}
    \mathcal{R}|i\rangle|0\rangle = |i\rangle \left[\sqrt{1 - p(i)}|0\rangle - \sqrt{p(i)}|1\rangle \right] = |\chi\rangle,
\end{equation}
We can also rewrite this as $|\chi\rangle$ as $\mathcal{F}|0^{N + 1}\rangle = \mathcal{R}\left(\mathcal{A} \otimes \mathcal{I}_{2} \right)|0^{N + 1}\rangle = |\chi\rangle$, where $\mathcal{F} = \mathcal{A}\mathcal{R}$. Furthermore, we can calculate the expectation value of state $|i\rangle |1\rangle$ as:
\begin{align}
    \mathbb{E}[p] &= \langle\chi |\left( \mathcal{I}_{N \times N}  \otimes |1\rangle \langle 1| \right)|\chi\rangle
    \\
    &= \sum_{i = 0}^{N-1}|\alpha_i|^2p_i,\label{eqn:expect_value}
\end{align}

To calculate the expected pairs produced, we need to embed the cross sections into the quantum circuit via the function $f(x)$. However, to this date, there is no simple way to embed a function as complicated as equation~\ref{eq:cross_section}; the most accurate solution would require quantum arithmetic, which would make the circuit too large for current NISQ devices. However, we can embed a polynomial approximation of said function into the quantum circuit. Stating the fact that 
$\sqrt{1 - f(x)} |0\rangle + \sqrt{f(x)}|1\rangle = \cos(\zeta(x))|0\rangle + \sin(\zeta(x)) |1\rangle$ for a given polynomial $\zeta(x) = \sum_{k=0}^l \zeta_kx^k$ of degree $l$. We can then find a polynomial which approximates equation~\ref{eq:cross_section} and embed it using controlled $Y$-rotation gates, $R_Y(\theta_i)$, where the angle $\theta_i = 2\sin^{-1}(\sqrt{p_i})$ . 
The theoretical probabilities for the range of energies have to be calculated before embedding them into the quantum circuit. Then the function must be approximated as a polynomial and implemented in the quantum circuit. The function then has to be scaled $f(x) \in [0,1]$ as follows: 
\begin{equation}
\tilde{f}(x) = \frac{f(x) - f_\textrm{min}}{f_\mathrm{max} - f_\mathrm{min}}.
\label{eqn:scale}    
\end{equation}
In this case, we use \verb|SciPy.optimize.curve_fit| \cite{virtanen_scipy_2020} to do the fitting of the function to a low-order polynomial function. Later, we can use the method \verb|PolynomialPauliRotations| from Qiskit Algorithms \cite{javadi-abhari_quantum_2024}, which can embed a polynomial of $\zeta$ as the angles into the $Y$-rotation gates.
After embedding the polynomial, the circuit is readily prepared to amplify the amplitude of the ``good" state in the ancilla qubit. To perform so, we use Qiskit Algorithms' \verb|IterativeAmplitudeEstimation|, which efficiently conducts the amplitude estimation using the algorithm introduced by Grinko \textit{et al.} \cite{grinko_iterative_2021} described earlier in this section.

To translate the expected value to the number of pairs, we need to solve for $f(x)$ in the equation \ref{eqn:scale}, and then we multiply by the number of interacting particles. 

In Figure~\ref{fig:algorithm pipeline}, We have the sketch of the algorithm denoting the principal steps taken: initialisation of the probability distribution using the three proposed methods, embedding of the cross sections as a polynomial approximation into the ancilla via controlled rotations \cite{woerner_quantum_2019}, use IQAE to obtain the expected probabilities, and post-process to obtain the expected number of pairs. 

Theoretically, IQAE could provide a quadratic speedup. However, in our work, this setup does show it. This is due to the error introduced by approximating the linear Breit-Wheeler cross section with a polynomial. If we were to use quantum arithmetic, we could obtain a quadratic speedup, but that would incur a circuit too deep for the NISQ era. Furthermore, the error that we would obtain in total would be $\epsilon_\text{tot} = \epsilon_\text{IQAE} + \epsilon_\text{approx}$

\bmhead{Acknowledgements}
L.I.I.G. would like to acknowledge Dr Bernardo Malaca, Lucas Ansia Fernandez and Dr Julien Zylberman for their insightful discussions on using the Deucalion Supercomputer, Monte Carlo methods, and state initialisation, respectively. We would also like to thank Dr. Vadim Karpusenko and IonQ for providing research credits to perform simulations on the IonQ cloud. This work was supported by the Portuguese Science Foundation (FCT) Grant No. 2023.16184.ICDT. This work has been carried out within the framework of the EUROfusion Consortium, funded by the European Union via the Euratom Research and Training Programme (Grant Agreement No. 101052200 – EUROfusion). Views and opinions expressed are, however, those of the authors only and do not necessarily reflect those of the European Union or the European Commission. Neither the European Union nor the European Commission can be held responsible for them.
We dedicate this work to the memory of Prof. Nuno Loureiro, for the support and the inspiration he shared with everyone around him.

\bibliography{QuantumMonteCarloHEP}

\end{document}